\documentclass[12pt]{article}
\def\ypiz_2gg{$\pi^0 \rightarrow \gamma\gamma$}
\newcommand {\charex} {$K^+\mathrm{Xe} \rightarrow K^0 p \mathrm{Xe}'$}
\newcommand {\under} {$K^+n \rightarrow K^0 p$}
\newcommand {\dimass} {$m(pK^0)$}

\begin {document}

\title
{Observation of a narrow baryon resonance with
positive strangeness formed in $K^+$Xe collisions}

\author{
DIANA Collaboration\\
V.V. Barmin$^a$, 
A.E. Asratyan$^{a,}$\thanks{Corresponding author. E-mail address:
asratyan@itep.ru.},
V.S. Borisov$^a$, 
C. Curceanu$^b$, \\
G.V. Davidenko$^a$, 
A.G. Dolgolenko$^a$,
C. Guaraldo$^b$, 
M.A. Kubantsev$^a$, \\
I.F. Larin$^a$, 
V.A. Matveev$^a$, 
V.A. Shebanov$^a$, 
N.N. Shishov$^a$, \\
L.I. Sokolov$^a$,
V.V. Tarasov$^a$,
G.K. Tumanov$^a$,
and V.S. Verebryusov$^a$ \\
{\normalsize $^a$ \it Institute of Theoretical and Experimental Physics,
Moscow 117218, Russia}\\
{\normalsize $^b$ \it Laboratori Nazionali di Frascati dell' INFN,
C.P. 13-I-00044 Frascati, Italy}
}                                          
\date {\today}
\maketitle

\begin{abstract}
The charge-exchange reaction \charex\ is investigated using the data of 
the DIANA experiment. The distribution of the $pK^0$ effective mass 
shows a prominent enhancement near 1538 MeV formed by nearly 80 events
above the background, whose width is consistent with being entirely due 
to the experimental resolution. Under the selections based on a 
simulation of $K^+$Xe collisions, the statistical significance of the 
signal reaches 5.5$\sigma$.
We interpret this observation as strong evidence for formation of a 
pentaquark baryon with positive strangeness, $\Theta^+(uudd\bar{s})$,
in the charge-exchange reaction \under\ on a bound neutron. The mass 
of the $\Theta^+$ baryon is measured as $m(\Theta^+) = 1538\pm2$ MeV.
Using the ratio between the numbers of resonant and non-resonant 
charge-exchange events in the peak region, the intrinsic width of 
this baryon resonance is determined as
$\Gamma(\Theta^+) = 0.34\pm0.10$ MeV.
\end{abstract}

\section{Introduction}

     The baryons built of four quarks and an antiquark as the lowest 
Fock component, referred to as pentaquarks, are not forbidden by theory 
and have been discussed ever since the appearance of the quark model 
\cite{forerunners}.
The critical prediction by Diakonov, Petrov and 
Polyakov \cite{DPP} has been that the lightest explicitly exotic 
baryon with positive strangeness, the $\Theta^+(uudd\bar{s})$, must 
be relatively light and narrow, which would have made its experimental
observation rather difficult.
Specifically, they predicted the mass $m \approx 1530$ MeV
and width $\Gamma < 15$ MeV for the $\Theta^+$, the lightest member of 
the pentaquark antidecuplet, that should decay to the $nK^+$ and $pK^0$ 
final states. More recent theoretical analyses suggest that the 
$\Theta^+$ intrinsic width may be on the order of 1 MeV or even 
less \cite{width}. 

     Narrow peaks near 1540 MeV 
in the $nK^+$ and $pK^0$ mass spectra were initially detected in
low-energy photoproduction by LEPS \cite{Nakano-old} and in the 
charge-exchange reaction $K^+n \rightarrow pK^0$ by DIANA 
\cite{DIANA-2003}. Subsequently, both experiments were able to confirm 
their initial observations \cite{Nakano,DIANA-2007,DIANA-2010}.
Moreover, increasing the statistics of the charge-exchange reaction 
allowed DIANA to directly estimate the $\Theta^+$ intrinsic width: 
$\Gamma \simeq 0.4\pm0.1$ MeV \cite{DIANA-2007,DIANA-2010}.
More recently, observation of a narrow peak near 1.54 GeV in the
missing mass of the $K^0_S$ meson in the reaction
$\gamma p \rightarrow K^0_S K^0_L p$ was reported by a group from
the CLAS collaboration \cite{Amaryan}.
Other searches for the $\Theta^+$ baryon in different reactions and
experimental conditions yielded both positive and negative results,
see the review papers \cite{Burkert} and \cite{Danilov-Mizuk} and
references therein. The bulk of null results can be probably explained
by the extreme smallness of the $\Theta^+$ width that implies the 
smallness of production cross-sections \cite{Diakonov}. 
Azimov {\it et al.} \cite{AziGoStra} argue that the published null 
results fail to rule out the existing positive evidence, and advocate 
a new approach to detecting the $\Theta^+$ in hard collisions.

     The charge-exchange reaction \under\ on a bound neutron, that 
is investigated by DIANA and BELLE \cite{BELLE}, is particularly 
interesting because it allows to probe the $\Theta^+$ intrinsic width 
in a model-independent manner. The existing data on low-energy $K^+ d$
scattering have been found to leave room for a $pK^0$ resonance with
mass near 1540 MeV, provided that its width is less than 1 MeV
\cite{Strakovsky,Cahn-Trilling,Sibirtsev-width,Gibbs,Azimov}.
An important advantage of the reaction \under\ is that the 
strangeness of the final-state $pK^0_S$ system is {\it a priori} known 
to be positive. In this paper, the DIANA data on the charge-exchange 
reaction \charex\ are analyzed using nearly 2.5 times more statistics 
than in \cite{DIANA-2003}.

\section{The experiment and the data}

     The DIANA bubble chamber \cite{chamber} filled with liquid Xenon was
exposed to a separated $K^+$ beam with momentum of 850 MeV from the
10-GeV proton synchrotron at ITEP, Moscow. The density and radiation 
length of the fill were 2.2 g/cm$^3$ and 3.7 cm, respectively. The 
chamber had a fiducial volume of $70\times70\times140$ cm$^3$ viewed by 
four optical cameras, and operated without magnetic field. 
In the fiducial volume of the bubble chamber, $K^+$ momentum varies 
from $\simeq730$ MeV for entering kaons to zero for those that range
out through ionization. Throughout this momentum interval, all collisions
and decays of incident $K^+$ mesons are efficiently detected. 
The $K^+$ momentum at interaction point is determined from the spatial 
distance between the detected vertex and the mean position of the 
vertices due to decays of stopping $K^+$ mesons. 
The estimate of the $K^+$ momentum based on the measured position of the 
interaction vertex has been verified by detecting and reconstructing the 
$K^+ \rightarrow \pi^+\pi^+\pi^-$ decays in flight, which provided an 
independent estimate of the $K^+$ momentum. Charged secondaries are 
identified by ionization and momentum-analyzed by their range in Xenon.
The detection efficiency for $\gamma$-quanta with $p_\gamma > 25$ MeV
is close to 100\%.

     On total, some $10^6$ tracks of incident $K^+$  mesons have been 
recorded on film. Scanning of the film yielded  nearly 55,000 events 
with visible $K^0_S$ decays, $K^0_S \rightarrow \pi^+\pi^-$ and 
$K^0_S \rightarrow \pi^0\pi^0$, inside the fiducial volume of the
bubble chamber. The ratio between the numbers of detected 
$K^0_S \rightarrow \pi^+\pi^-$ and $K^0_S \rightarrow \pi^0\pi^0$
decays is consistent with the ratio between the corresponding $K^0_S$
branching fractions \cite{PDG}.
These $K^0_S$ decays could be associated with primary 
$K^+$Xe vertices with various multiplicities of secondary particles. 
Finally, events that featured a $K^0_S \rightarrow \pi^+\pi^-$ decay,
a measurable proton with track length over some 3.5 mm, and no 
additional measurable or stub-like protons in the final state, were 
selected as candidates for the charge-exchange reaction \under\ free of
intranuclear rescatterings. The $K^0_S \rightarrow \pi^+\pi^-$ decays
with a large spatial angle between the decay pions, 
$\Theta_{\pi\pi} > 150^{0}$, were rejected.

     The selected events are then fully 
measured and reconstructed in space using specially designed 
stereo-projectors. In a selected event, we measure the  
polar and azimuthal angles of the $K^0_S$ and proton 
with respect to the $K^+$ direction, similar angles of the
$\pi^+$ and $\pi^-$ with respect to the parent $K^0_S$
direction, and the proton and pion ranges in Xenon. We additionally
measure the opening angle between the $K^0_S$ and proton directions
which allows the most accurate estimate of the $pK^0_S$ effective mass.
The momentum is estimated by range for the proton, and by kinematic
reconstruction for the $K^0_S$ using the ranges and emission angles of
decay pions. For further 
rejection of $K^0_S$ mesons that may have scattered by small angles in 
liquid Xenon but passed the pointback criteria, we then apply the 
selection $\tau < 3\tau_0$ where $\tau$ is the $K^0_S$ measured proper 
lifetime and $\tau_0$ is its tabulated mean value \cite{PDG}.
The quality of the data is best reflected by the experimental 
resolution on effective mass of the $pK^0_S$ system, estimated as 
$\sigma_m \simeq 3.5$ MeV by error propagation for observed events
and by a simulation.
As expected, the resolution on the $pK^0_S$ effective mass is similar 
to the instrumental width of the $\Lambda \rightarrow p \pi^-$ peak, 
$\sigma = 3.3\pm1.0$ MeV, previously observed in the antiproton 
exposure of DIANA \cite{lambda,DIANA-2003}.
Further details on the experimental procedure may be found in
\cite{exp-procedure-1,exp-procedure-2,DIANA-2007}.
\begin{figure}[!t]

\vspace{4.5 cm}
\includegraphics{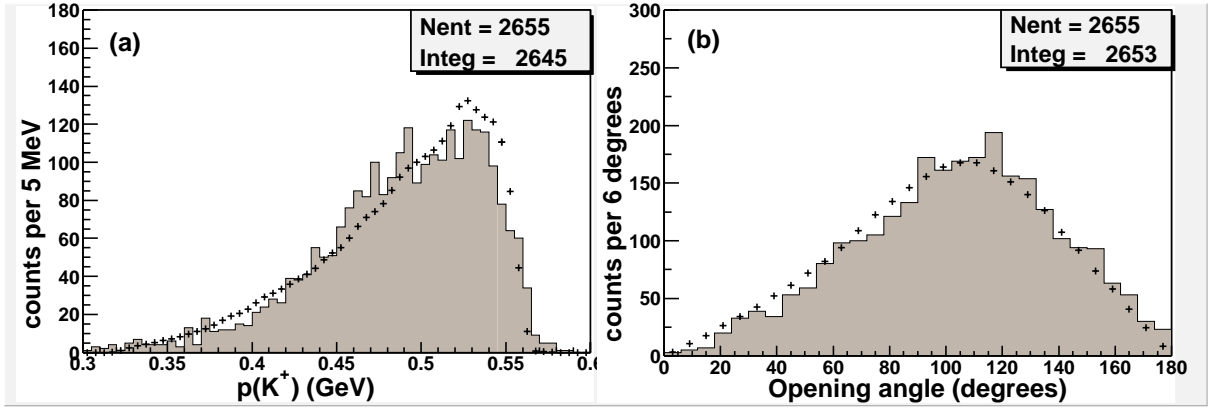}
\caption {(Color online)
Incident $K^+$ momentum at collision point (a) and the opening angle 
between the $K^0_S$ and proton directions in lab (b). The crosses show 
the simulated distributions that have been normalized to the number of 
live events (see Section 3).}
\label{pbeam}
\end{figure}

     The measurements have been restricted to the region
$L(K^+) > 520$ mm, where $L(K^+)$ is the length of the $K^+$ path in 
liquid Xenon before the collision. (Note that there is no one-to-one 
correspondence between $L(K^+)$ and $K^+$ momentum, because the original
beam momentum varied by some $\pm20$ MeV in different exposures.)
The laboratory momentum of the incident $K^+$ at collision point
is shown in Fig.~\ref{pbeam}a for all measured events of the reaction
\charex\ with $K^0_S$ and proton momenta above 155 and 165 MeV, 
respectively (instrumental thresholds). The measured opening angle
between the $K^0_S$ and proton directions is shown in Fig.~\ref{pbeam}b.
The dataset comprises the data treated in our initial analysis
\cite{DIANA-2003} and the subsequent measurements. The statistics of 
the charge-exchange reaction has been increased by a factor $\simeq2.5$ 
as compared to \cite{DIANA-2003}.

\section { The Monte-Carlo description of the data }

     Rescattering of either the $K^0$ or proton in the Xenon nucleus
distorts the effective mass of the $pK^0$ system originally formed
in the charge-exchange reaction \under\ on a bound neutron.
In formulating the selection criteria for unrescattered events, we 
rely on a Monte-Carlo simulation of $K^+n$ and $K^+p$ collisions in
nuclear medium. We simulate the original collision that may be either
\under, $K^+n \rightarrow K^+n$, or $K^+p \rightarrow K^+p$, and then
follow the development of the intranuclear cascade that also involves
the elastic NN reactions $np \rightarrow np$, $nn \rightarrow nn$, 
$pn \rightarrow pn$, and $pp \rightarrow pp$. In order to reproduce
the experimental selections for the measured events, we then select 
those simulated events that feature a final-state 
$K^0 \rightarrow \pi^+\pi^-$  with lab momentum $p_K > 155$ MeV
and $\Theta_{\pi\pi} < 150^{0}$,
a proton with $p_p > 165$ MeV, and no extra 
protons with $p_p > 100$ MeV which corresponds to the experimental 
threshold for proton detection. On the other hand, any number of 
emitted neutrons is allowed.

\begin{figure}[!t]

\vspace{9.9cm}
\includegraphics{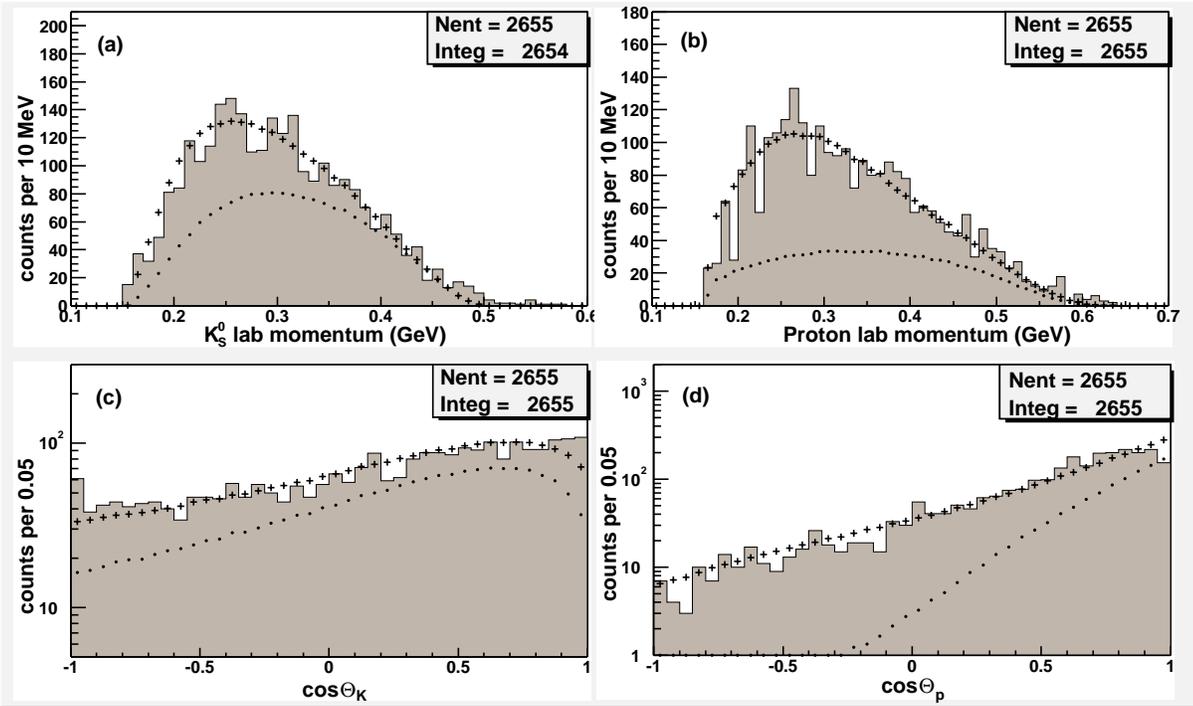}
\caption {(Color online)
Laboratory momenta of the $K^0$ (a) and proton (b) and the cosines
of the $K^0$ (c) and proton (d) emission angles with respect to the
beam. The crosses show the corresponding distributions of all simulated 
events that have been normalized to the number of measured events. 
Additionally shown by dots in (a) and (c) are the simulated spectra of 
unrescattered $K^0$ mesons, and in (b) and (d) --- of unrescattered 
protons.}
\label{momenta}
\end{figure}

     The cross-sections of the aforementioned $KN$ and $NN$ reactions
as functions of collision energy are parametrized using the existing
data \cite{cross-section,Dover,hadronic-xsections}. We substitute
$\sigma(K^+n\rightarrow K^+n) = \sigma(K^+d\rightarrow K^+d) -
\sigma(K^+p\rightarrow K^+p)$, and invoke the isospin relations 
$\sigma(K^0n\rightarrow K^0n) = \sigma(K^+p\rightarrow K^+p)$ and
$\sigma(K^0p\rightarrow K^0p) = \sigma(K^+n\rightarrow K^+n)$ for 
the $K^0N$ elastic cross sections that have not been measured
directly. The effective radius 
of the target nucleus is taken in the form $r = 1.25 \times A^{1/3}$ fm 
where $A = 131$ for Xenon, and the neutron and proton densities are 
assumed to be uniform throughout the nucleus volume. For the same nucleus, 
we use a realistic form of the Fermi-momentum distribution with maximum 
near 160 MeV \cite{Zhalov}. For the unbound nucleons, Pauli blocking is 
approximated by the cut $p_N > 180$ MeV on nucleon momentum, and
absorption is treated according to \cite{Bernardini}. 
For the real intranuclear potentials of the nucleon and the $K^+$ meson
in the Xenon mucleus, we assume $V_N = -40$ MeV and $V_K = +25$ MeV 
\cite{Dover,kaon-potential}. The flux of 
incident $K^+$ mesons as a function of $K^+$ momentum at collision point 
is inferred from the observed distribution of $K^+$ range in Xenon before 
interaction or decay, see \cite{DIANA-2003}. The experimental losses
of soft protons and $K^0_S$ mesons, that largely occur at lab momenta
below some 200 MeV, are accounted for. The experimental 
uncertainties and measurement errors are included in the simulation.
The simulation adequately reproduces the proportion among the numbers
of scanned events with different multiplicities of detected protons.
\begin{figure}[!t]

\vspace{9.9cm}
\includegraphics{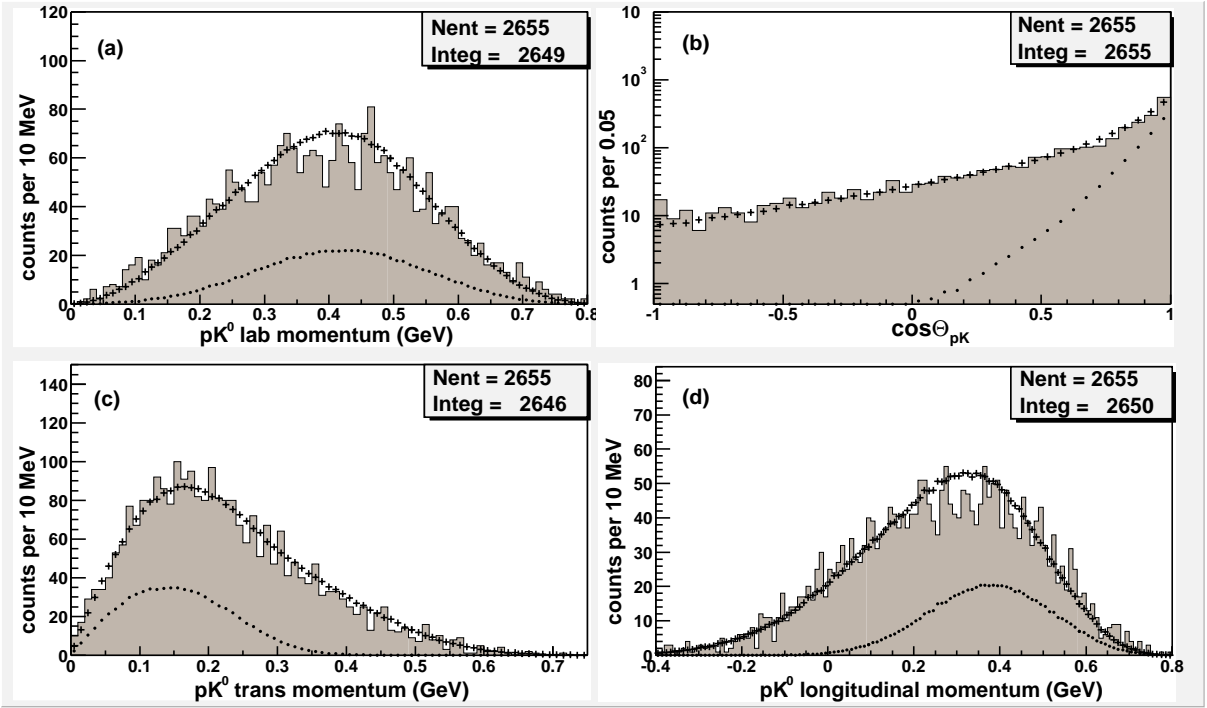}
\caption {(Color online)
The absolute lab momentum of the $pK^0$ system (a) ;
the cosine of the $pK^0$ emission angle in lab (b) ; and
the transverse (c) and longitudinal (d) momenta of the $pK^0$ system.
The crosses show the corresponding distributions of all 
simulated events that have been normalized to the number of live
events. Depicted by dots are the simulated distributions of
rescattering-free events.}
\label{others}
\end{figure}

     In Figures \ref{momenta} and \ref{others}, some distributions
of measured (or live) events are compared with those of simulated events.
Here and in what follows, the total number of simulated events is 
normalized to that of live events prior to analysis selections. 
Laboratory momenta of the $K^0$ and proton are shown in 
Figs.~\ref{momenta}a and \ref{momenta}b, and their emission angles 
with respect to the incident $K^+$ --- in Figs.~\ref{momenta}c and
\ref{momenta}d. Shown by dots in Figs.~\ref{momenta}a and 
\ref{momenta}c are the simulated spectra of unrescattered $K^0$
mesons (in the same event, the proton may rescatter or not).
Simularly, the dots in Figs.~\ref{momenta}b and \ref{momenta}d are 
the spectra of unrescattered protons (the $K^0$ may rescatter or
not). More originally-produced $K^0$ mesons than protons 
are seen to escape from the nucleus without rescattering.
On average, the rescattered $K^0$ mesons and protons have
smaller momenta and broader emission angles than the unrescattered
ones. Therefore, rejecting the $K^0$ mesons and protons that travel
in the backward hemisphere in lab will enhance the fraction of
rescattering-free events (those in which both products of the 
primary \under\ collision escaped from the nucleus without 
rescattering).

     The quantities that describe the $pK^0$ system as a whole
are plotted in Fig.~\ref{others}. Here, the dots refer to
the rescattering-free \under\ collisions. The distributions of 
rescattering-free and rescattered events have similar shapes for the 
absolute lab momentum of the $pK^0$ pair (Fig.~\ref{others}a),
but very different shapes for its transverse and longitudinal 
components $p_T$ and $p_L$, see Figs. \ref{others}c and \ref{others}d. 
The bulk of rescattering-free events lie in the region $p_T < 300$ MeV, 
whereas the rescattered events transcend the domain of target 
Fermi-motion by reaching up to some 600 MeV. On the other hand, the 
simulation predicts that rescattering-free events should populate the 
region $p_L > 100$ MeV unlike the rescattered ones that extend to 
negative values of $p_L$. As a result, the rescattered $pK^0$ systems 
are emitted at broader angles to the $K^+$ beam than the unrescattered 
ones, see Fig.~\ref{others}b. Therefore, the fraction of rescattered 
events will be reduced by rejecting events with large $p_T$ and small 
$p_L$ of the emitted $pK^0$ system, or those emitted at broad angles 
to the beam.

\section{The signal of the $\Theta^+$ baryon prior to analysis selections}

\begin{figure}[!b]

\vspace{15.5cm}
\includegraphics{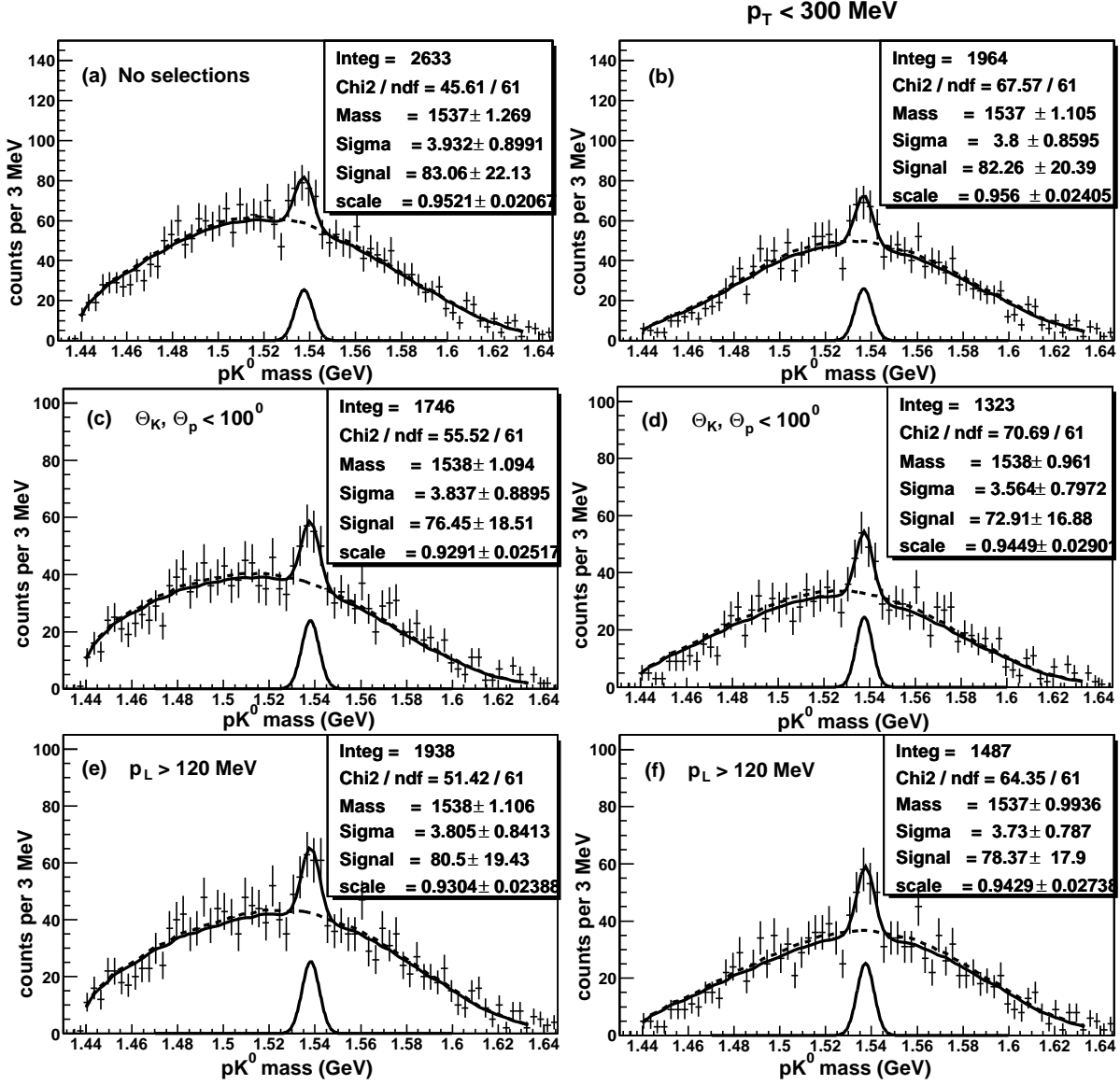}
\caption
{In (a), the original \dimass\ distribution is fitted to a Gaussian 
plus a background function, obtained by scaling the simulated 
distribution by a factor which is a free parameter of the fit. 
The dashed line shows the null fit to the background function alone.
Shown and fitted in (c) and (e) are the $pK^0$ effective-mass spectra 
under the selections $\Theta_K, \Theta_p < 100^0$ and $p_L > 120$ MeV,
respectively (see Section 5).
The selection $p_T < 300$ MeV is additionally applied in the 
right-hand panels (b), (d), and (f).}
\label{1dim-all-and-sele}
\end{figure}

     The $pK^0$ effective-mass spectrum for all measured events of
the reaction \charex, that is shown in Fig.~\ref{1dim-all-and-sele}a,
is enhanced in the region \dimass\ $\simeq 1538$ MeV. 
This distribution is fitted to a Gaussian plus a background
function, constructed by scaling the simulated \dimass\ distribution 
by a factor that is treated as a free parameter of the fit. (In this 
and subsequent fits, the maximum-likelihood algorithm is used.) 
The fitted position of the enhancement is close to 
1538 MeV, and its fitted width is consistent with the simulated 
experimental resolution on \dimass\ : $\sigma_m \simeq 3.5$ MeV. 
As compared with our initial analysis \cite{DIANA-2003}, the fitted 
signal has increased in magnitude according to the increase of the 
total statistics of measured events. The same distribution is 
also fitted to the background function alone, which corresponds to 
the null hypothesis. (This is shown by the dashed line in 
Fig.~\ref{1dim-all-and-sele}a.) The naive estimate of statistical
significance is $S/\sqrt{B} = 4.8\sigma$, where the signal $S$ and 
background $B$ have been derived from the signal hypothesis alone 
over the 90\% area of the Gaussian.

\begin{figure}[!t]

\vspace{12.5cm}
\includegraphics{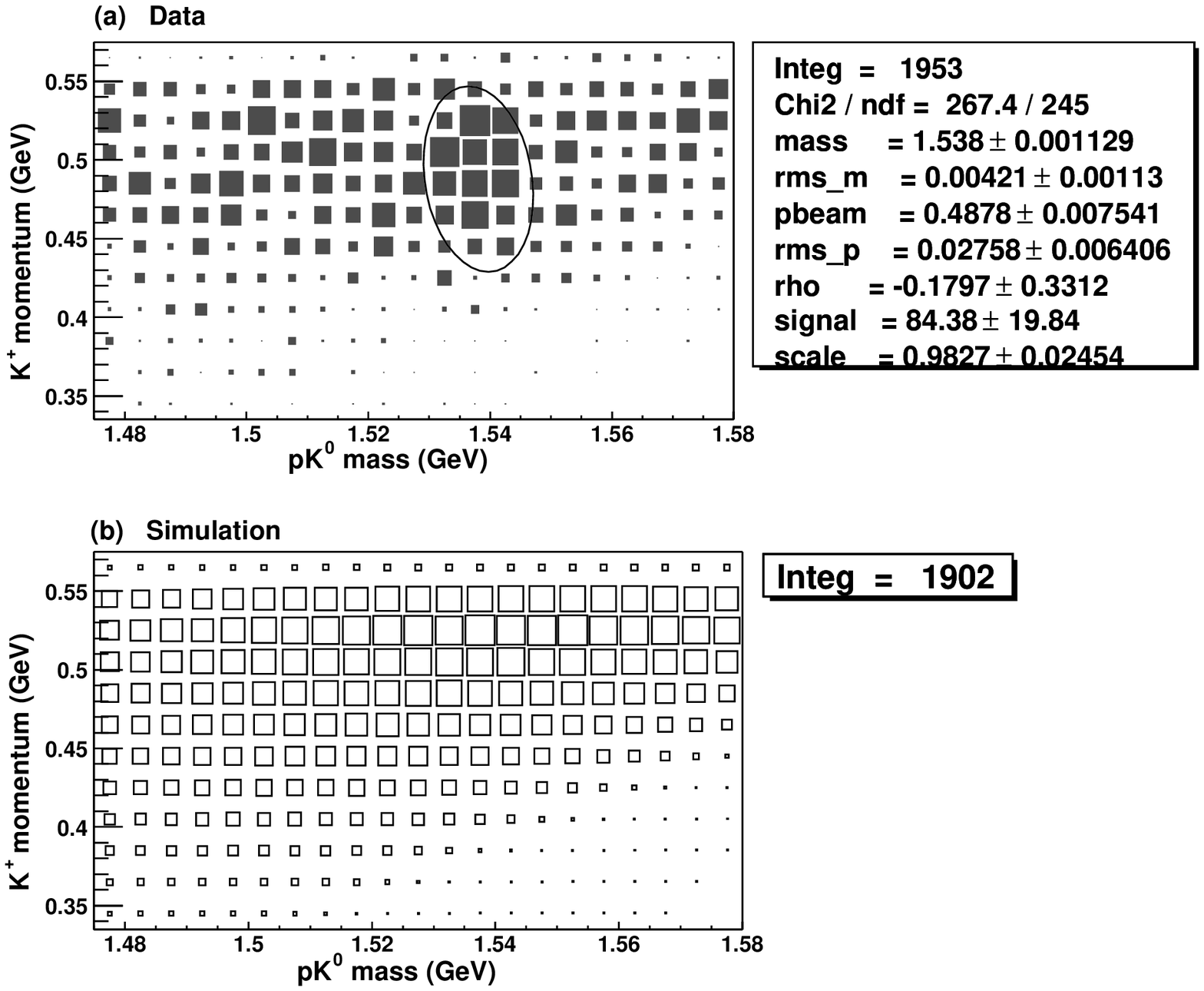}
\caption {(Color online)
The scatter plots in \dimass\ and $p(K^+)$ for all live (a) and 
simulated (b) events. Also shown in (a) is the fit to a two-dimensional 
Gaussian plus the background function. The ellipse in (a) is the 90\% 
contour for the two-dimensional Gaussian.}
\label{2dim-all}
\end{figure}

     For formation of the putative pentaquark baryon $\Theta^+(1540)$ 
in the reaction \under\ on a free stationary neutron, the resonant 
value of beam momentum $p(K^+)$ is $\simeq445$ MeV. For $\Theta^+$ 
formation on a bound neutron, $p(K^+)$ will be shifted up by some 50
MeV by the $K^+$ intranuclear potential, and smeared by Fermi motion of 
the neutron target. Despite the smearing of the resonance lineshape in
$p(K^+)$ by nuclear effects, a combined analysis of the two variables may 
prove to be more sensitive to $\Theta^+$ formation than the analysis of 
\dimass\ alone. The scatter plot in \dimass\ and $p(K^+)$ for all live 
events, shown in Fig.~\ref{2dim-all}a, is indeed enhanced in the region 
\dimass\ $\simeq 1540$ MeV and $p(K^+) \simeq 500$ MeV. The corresponding
scatter plot for simulated events proves to be regular over the full area 
of \dimass\ and $p(K^+)$, see Fig.~\ref{2dim-all}b.
In Fig.~\ref{2dim-all}a, the distribution of live events is fitted
to a two-dimensional Gaussian plus a background function, again
constructed by scaling the simulated distribution by a factor which
is a free parameter of the fit. The same distribution has also been
fitted to the background function alone, which corresponds to the 
null hypothesis. 

     The correlation parameter of the two-dimensional Gaussian (line 7 
in the box in Fig.~\ref{2dim-all}a) is consistent with $\rho = 0$, as 
physically expected for formation of a narrow $pK^0$ resonance.
The enhancement is centered at \dimass $\simeq 1538$ MeV and 
$p(K^+) \simeq 490$ MeV, see lines 3 and 5 in the box. The rms width
of the enhancement in \dimass\ is consistent with the experimental
resolution, and that in $p(K^+)$ is $\simeq 28$ MeV (lines 4 and 6 in 
the box). The observed spread of the signal in $p(K^+)$ is consistent 
with the smearing of a narrow $pK^0$ resonance by nuclear effects 
\cite{DIANA-2007}. The fitted signal (line 8) is in good agreement 
with the one-dimensional signal in Fig.~\ref{1dim-all-and-sele}a,
but proves to be more significant : $S/\sqrt{B} = 5.1$. This is 
because the fitted signal is spread over a narrower interval of 
$p(K^+)$ than the nonresonant background.

\section { Applying additional selections }

\begin{figure}[!b]

\vspace{13.5cm}
\includegraphics{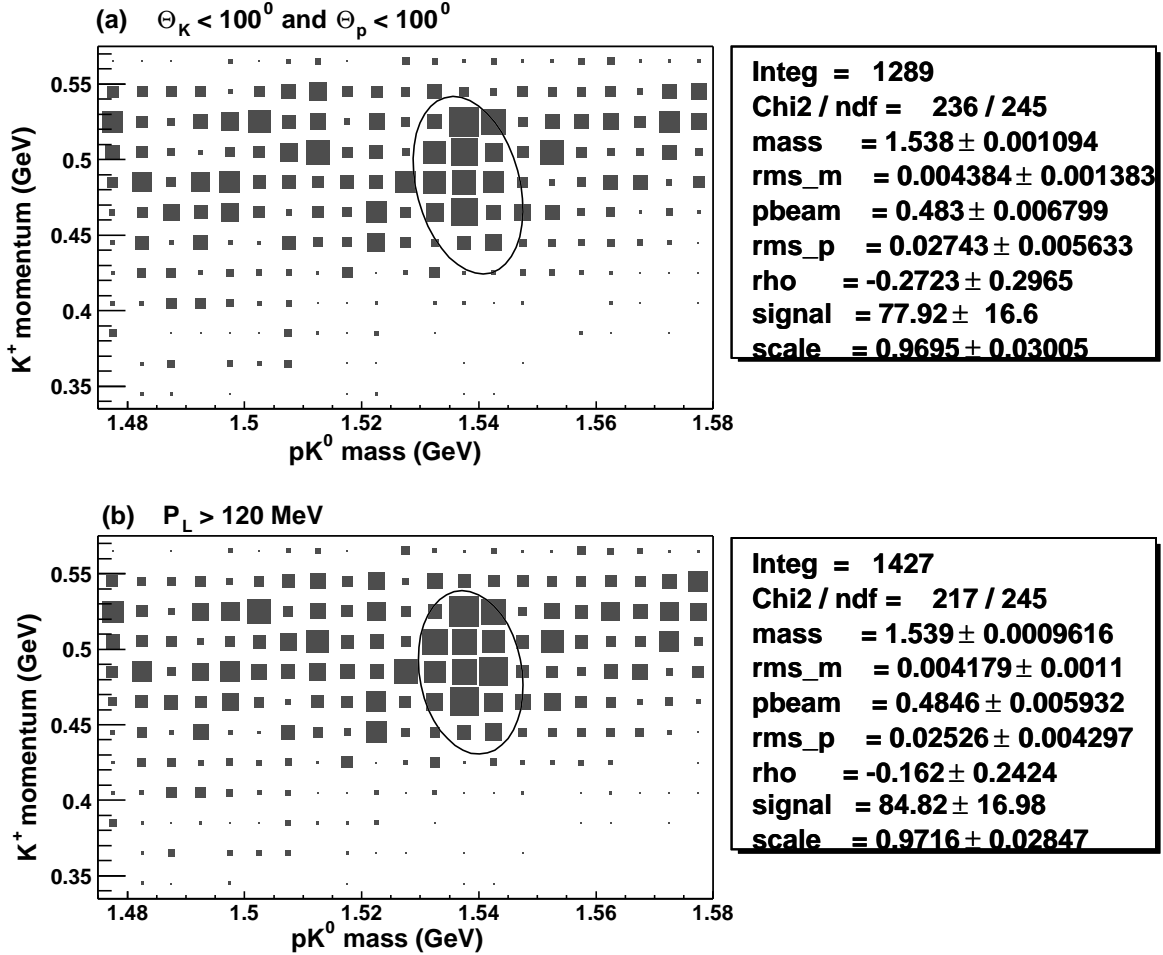}
\caption {(Color online)
Shown in (a) and (b) are the scatter plots in \dimass\ and $p(K^+)$ 
under the additional selections $\Theta_K, \Theta_p < 100^0$ and 
$p_L > 120$ MeV, respectively. Either scatter
plot is fitted to a two-dimensional Gaussian plus a background
function obtained by scaling the simulated distribution under
similar selections (not shown). The ellipses are the 90\% contours 
for the two-dimensional Gaussian.}
\label{2dim-sele}
\end{figure}

     In order to verify that the enhancement at \dimass\ $\simeq 1538$ 
MeV is formed by rescattering-free events, as expected for the signal
of a narrow $pK^0$ resonance, one has to use additional selections 
that reduce the fraction of rescattered events. We apply the following
selections:
\begin{enumerate}
\item 
$\Theta_K < 100^0$ and $\Theta_p < 100^0$ for the $K^0$ and proton
emission angles, suggested by the distributions of these variables 
shown in Figs.~\ref{momenta}c and \ref{momenta}d. This selection has 
already been used in our previous papers \cite{DIANA-2003,DIANA-2007,
DIANA-2010}, and was found to produce no artificial structures in the 
\dimass\ spectrum by an independent theoretical analysis \cite{Sibirtsev}. 
The simulation predicts that this selection retains 77\% of all
rescattering-free events.
\item
$p_L > 120$ MeV for the longitudinal lab momentum of the $pK^0$
system, as suggested by the data shown in Fig.~\ref{others}d.
The acceptance to simulated rescattering-free events is $\simeq96$\%.
\end{enumerate}

The effects of the selections  $\Theta_K, \Theta_p < 100^0$ and 
$p_L > 120$ MeV are shown in Figs. 
\ref{1dim-all-and-sele}c and \ref{1dim-all-and-sele}e, respectively.
Each mass spectrum is fitted to a Gaussian plus a background 
function, that is constructed by scaling the simulated distribution 
under similar selections. The signal and null fits are shown by the
solid and dashed lines, respectively. The value of $S/\sqrt{B}$ is
5.6 for the enhancement in Fig.~\ref{1dim-all-and-sele}c, and 5.5
for that in Fig.~\ref{1dim-all-and-sele}e. Each additional selection 
is seen to render the signal more significant than in 
Fig.~\ref{1dim-all-and-sele}a. The selection $p_T < 300$ MeV, that
is suggested by the data of Fig.~\ref{others}c, is additionally
applied in the right-hand panels of Fig.~\ref{1dim-all-and-sele}.
This further increases the $S/\sqrt{B}$ value by $\simeq 0.4$.

     The experimental scatter plots in \dimass\ and $p(K^+)$
under the selections $\Theta_K, \Theta_p < 100^0$ and $p_L > 120$ MeV
are shown and fitted in Figs. \ref{2dim-sele}a and \ref{2dim-sele}b, 
respectively. (The simulated scatter plots under these selections are 
again regular throughout the full area of \dimass\ and $p(K^+)$.)
The positions and rms widths of the enhancement are consistent
with those in Fig.~\ref{2dim-all}. The two-dimensional signals in 
Figs \ref{2dim-sele}a and \ref{2dim-sele}b are similar in magnitude 
to the corresponding one-dimensional signals in Figs. 
\ref{1dim-all-and-sele}c and \ref{1dim-all-and-sele}e, but have 
higher values of $S/\sqrt{B}$ (5.8 and 6.2, respectively).

\begin{figure}[t]

\vspace{5.1cm}
\includegraphics{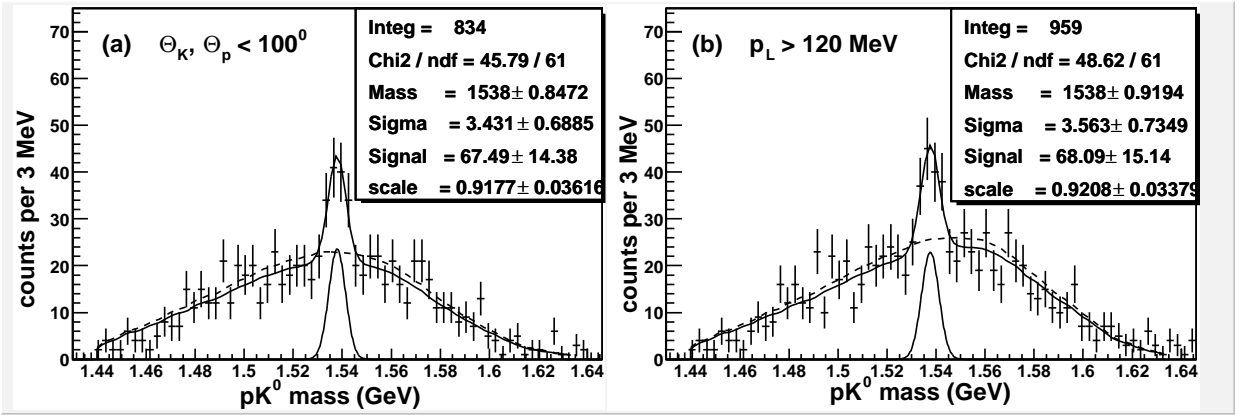}
\caption
{Shown in (a) and (b) are the $pK^0$ effective-mass spectra 
under the selections $\Theta_K, \Theta_p < 100^0$ and $p_L > 120$ MeV 
plus the common selections
$p_T < 300$ MeV and $445 < p(K^+) < 535$ MeV. The signal and
null fits are shown by the solid and dashed lines, respectively.}
\label{1dim-window-tra}
\end{figure}

     The fits of the scatter plots in \dimass\ and $p(K^+)$ 
shown in Figs. \ref{2dim-all} and \ref{2dim-sele}
suggest that the signal populates a limited range of $p(K^+)$, as it 
should for formation of a narrow $pK^0$ resonance \cite{DIANA-2007}. 
In Fig.~\ref{1dim-window-tra} the $pK^0$ effective mass is plotted
under the selections $\Theta_K, \Theta_p < 100^0$ or $p_L > 120$ MeV
plus the common selections $p_T < 300$ MeV and $445 < p(K^+) < 535$ MeV. 
The null fits demonstrate that the extra selection $445 < p(K^+) < 535$ 
MeV does not produce any spurious structures in the \dimass\ spectrum, 
while substantially increasing the signal-to-background ratio:
we have $S/\sqrt{B} = 6.8$ and 6.4 for the signals in Figs. 
\ref{1dim-window-tra}a and \ref{1dim-window-tra}b, respectively.

\section{Statistical significance of the signal}

      In all one- and two-dimensional fits shown in the previous
sections, the fitted width of the enhancement in \dimass\ is consistent 
with being entirely due to the experimental resolution. So in order to 
reduce the number of free parameters, the mass width is constrained to
the simulated value of $\sigma_m = 3.5$ MeV when estimating the 
statistical significance of the signal. 

     For the fits of the scatter plots in \dimass\ and $p(K^+)$ shown
in Figs. \ref{2dim-all} and \ref{2dim-sele}, the correlation parameter 
of the two-dimensional Gaussian is always consistent with $\rho = 0$, 
as physically expected for the signal from formation of a narrow $pK^0$ 
resonance. Indeed, in this case the variation of \dimass\ is totally 
due to measurement errors on the $K^0$ and proton momenta and on the 
opening angle, and should be fully independent from the variation of 
$p(K^+)$ that arises from Fermi motion of the target neutron. Therefore,
we use the constraint $\rho = 0$ when estimating the statistical 
significance of the two-dimensional signal.

     The results of the constrained fits of the $pK^0$ mass spectra 
and of the scatter plots in \dimass\ and $p(K^+)$ are shown in Tables 
\ref{constrained-1dim} and \ref{constrained-2dim}, respectively.
Also shown for each fit is the difference between the log-likelihood
values for the signal and null hypotheses, $-2\Delta\ln L$. 
The number of degrees of freedom is $\Delta\mathrm{ndf} = 2$ and 4 
for the constrained one- and two-dimensional fits, respectively. 
The statistical significance of the signal is estimated using the
value of $\chi^2$ for one degree of freedom which corresponds to
the same $p$-value as $\chi^2 = -2\Delta\ln L$ for $\Delta$ndf
degrees of freedom \cite{PDG}.

     We see that the statistical significance of the signal is 
enhanced by the additional kinematic selections based on the Monte-Carlo
simulation, reaching some 5.5 standard deviations. The two-dimensional
signals are more significant than the one-dimensional ones under
similar selections.
\clearpage

\begin{table}
\small
\begin{tabular}{|l|l|l|l|l|c|c|}
\hline
Selections & $m_0$ (MeV) & Signal (ev)
& $-2 \ln L$ & $-2 \ln L$ & $2\Delta\ln L$   & Stat. \\
        &         &  $S/\sqrt{B}$
& $\chi^2/$ndf & $\chi^2/$ndf & & sign. \\
            &                 &
& (signal fit)    & (null fit)        &         &                \\
\hline
\hline
None                & $1537\pm1$ & $77.8\pm18.8$
& 47.6   & 65.7   & 18.1 &  4.0                            \\
                     &             &  4.8
& 45.9/62   & 62.9/64   & &\\
\hline
$p_T < 300$ MeV     & $1537\pm1$ & $79.1\pm18.0$
& 65.7   & 88.9   & 23.2 & 4.4                            \\
                     &             &  5.3
& 67.8/62   & 91.9/64   & &\\
\hline
$p_T < 300$ MeV     & $1537\pm1$ & $67.7\pm15.0$
& 51.6   & 75.0   & 23.4 & 4.4                            \\
$445 < p(K^+) < 535$ MeV       &            & 5.6 
& 45.7/62   & 66.2/64   &  &\\
\hline
\hline
$\Theta_K,\Theta_p<100^0$ & $1538\pm1$ & $72.9\pm15.8$
& 69.0   & 92.8   & 23.8 & 4.5                              \\
                     &             & 5.6
& 55.7/62   & 77.0/64   & &\\
\hline
$\Theta_K,\Theta_p<100^0$      & $1538\pm1$ & $72.3\pm15.2$
& 80.9   & 109.9   & 28.9 & 5.1                               \\
$p_T < 300$ MeV     &             & 6.0
& 70.7/62   & 99.8/64   & &\\
\hline
$\Theta_K,\Theta_p<100^0$      & $1538\pm1$ & $68.0\pm13.4$
& 61.8   & 96.9   & 35.1 & 5.5                               \\
$p_T < 300$ MeV     &             & 6.8
& 45.8/62   & 77.3/64   & &\\
$445 < p(K^+) < 535$ MeV                    &             &  
&     &     &     &                                 \\
\hline
\hline
$p_L > 120$ MeV     & $1538\pm1$ & $77.1\pm16.7$
& 58.0   & 81.8   & 23.8 & 4.5                               \\
                     &             & 5.5
& 51.6/62   & 72.8/64   & &\\
\hline
$p_L > 120$ MeV  & $1537\pm1$ & $76.1\pm16.0$
& 68.6   & 97.1   & 28.5 & 5.0                               \\
$p_T<300$ MeV   &             &  6.0
& 64.5/62   & 93.0/64   & &\\
\hline
$p_L > 120$ MeV  & $1538\pm1$ & $67.6\pm13.9$
& 57.8   & 89.1   & 31.3 & 5.3                               \\
$p_T<300$ MeV   &             &  6.4
& 48.6/62   & 78.5/64   & &\\
$445 < p(K^+) < 535$ MeV                    &             &  
&     &     &     &                                 \\
\hline
\end{tabular}

\caption
{The results of the one-dimensional fits in which the Gaussian mass 
width of the signal has been constrained to the simulated resolution 
of $\sigma_m = 3.5$ MeV.}
\label{constrained-1dim}
\end{table}

\begin{table}
\small
\begin{tabular}{|l|l|l|l|l|l|c|c|}
\hline
Selections & $m_0$ (MeV)  & $p_0(K^+)$ (MeV) & Signal (ev)
& $-2 \ln L$ & $-2 \ln L$ &$2\Delta\ln L$ & Stat. \\
         &          & $\sigma_p$ (MeV)        &  $S/\sqrt{B}$
& $\chi^2/$ndf & $\chi^2/$ndf & & sign. \\
         &            &                 &
& (signal fit)    & (null fit)        &         &                \\
\hline
\hline
None                & $1538\pm1$ & $488.2\pm7.3$ & $79.4\pm18.7$
& 315.2 & 341.8 & 26.6 & 4.3                                \\
                    &         & $26.3\pm6.0$ & 5.4
& 268.5/247 & 305.8/251 & &\\
\hline
$\Theta_K,\Theta_p<100^0$ & $1538\pm1$ & $484.9\pm6.4$ & $73.1\pm15.5$
& 315.8 & 350.2 & 34.3 & 5.0                                 \\
                           &        & $25.8\pm4.7$ & 6.1
& 236.7/247 & 279.7/251 & &\\
\hline
$p_L > 120$ MeV     & $1539\pm1$ & $485.3\pm6.0$ & $81.0\pm16.3$
& 295.6 & 333.9 & 38.2 & 5.3                                 \\
                     &        & $24.5\pm4.2$ & 6.5
& 218.0/247 & 266.0/251 & &\\
\hline
\end{tabular}

\caption
{The results of the two-dimensional fits in which the Gaussian mass 
width of the signal has been constrained to the simulated resolution 
of $\sigma_m = 3.5$ MeV and the correlation parameter $\rho$ has 
been constrained to zero.}
\label{constrained-2dim}
\end{table}

\section{Intrinsic width of the $\Theta^+$ baryon}

     Intrinsic width of a resonance formed in an $s$-channel reaction
like \under\ can be estimated by comparing the signal magnitude with
the level of non-resonant background under the peak, see {\it e.g.}
\cite{Cahn-Trilling}. However, this method cannot be directly applied
to $K^+$ collisions with heavy nuclei because the resonant signal and 
the underlying non-resonant background may be very differently affected
by rescattering of the $K^0$ and proton in nuclear medium. That is, the 
non-resonant background under the peak is a mixture of unrescattered
and rescattered events, whereas a true signal should consist of 
unrescattered events only. 

     As soon as the $\Theta^+$ decay width is on the order of
1 MeV or less, the peak will not be depleted by the $K^0$
and proton rescatterings because the bulk of produced $\Theta^+$ 
baryons will decay upon leaving the nucleus. Therefore, for a 
consistent determination of the $\Theta^+$ intrinsic width based on
the method \cite{Cahn-Trilling} one will need the distribution of
non-resonant events in the effective mass of the originally formed 
$pK^0$ system prior to any rescatterings, $m_0(pK^0)$.
The $m_0(pK^0)$ distribution of non-resonant events can only be 
obtained through a simulation. Pauli blocking for protons in nuclear 
matter does not affect the process of $\Theta^+$ formation
and decay, and therefore should be switched off when consistently
simulating the ``equivalent" non-resonant background.
Then, assuming $J = 1/2$ for the $\Theta^+$ spin and using the 
observed signal and simulated non-resonant background, the 
intrinsic width of the $\Theta^+$ baryon can be derived as
\begin{equation}
\Gamma = \frac{N_\mathrm{peak}}{N_\mathrm{bkgd}}
\times \frac{\sigma^\mathrm{CE}}{107\mathrm{mb}}
\times \frac{\Delta m_0}{B_i B_f}.
\end{equation}
Here, $N_\mathrm{peak}$ is the fitted number of events in the peak
\vspace{.1in}
corrected for experimental losses; \\
$\Delta m_0$ is the interval of the original mass $m_0(pK^0)$
centered on peak position, that is populated by $N_\mathrm{bkgd}$ 
\vspace{.1in}
simulated non-resonant events; \\
$\sigma^\mathrm{CE} = 4.1\pm0.3$ mb is the measured cross section  
of the charge-exchange reaction \under\ for the center-of-mass energy
\vspace{.1in}
equal to the $\Theta^+$ mass \cite{cross-section}; \\
and $B_i$ and $B_f$ are the branching fractions for the initial and 
final states ($B_i = B_f = 1/2$ for the $\Theta^+$ isospin of either 
$I = 0$ or $I = 1$).
\begin{figure}[h]

\vspace{6cm}
\includegraphics{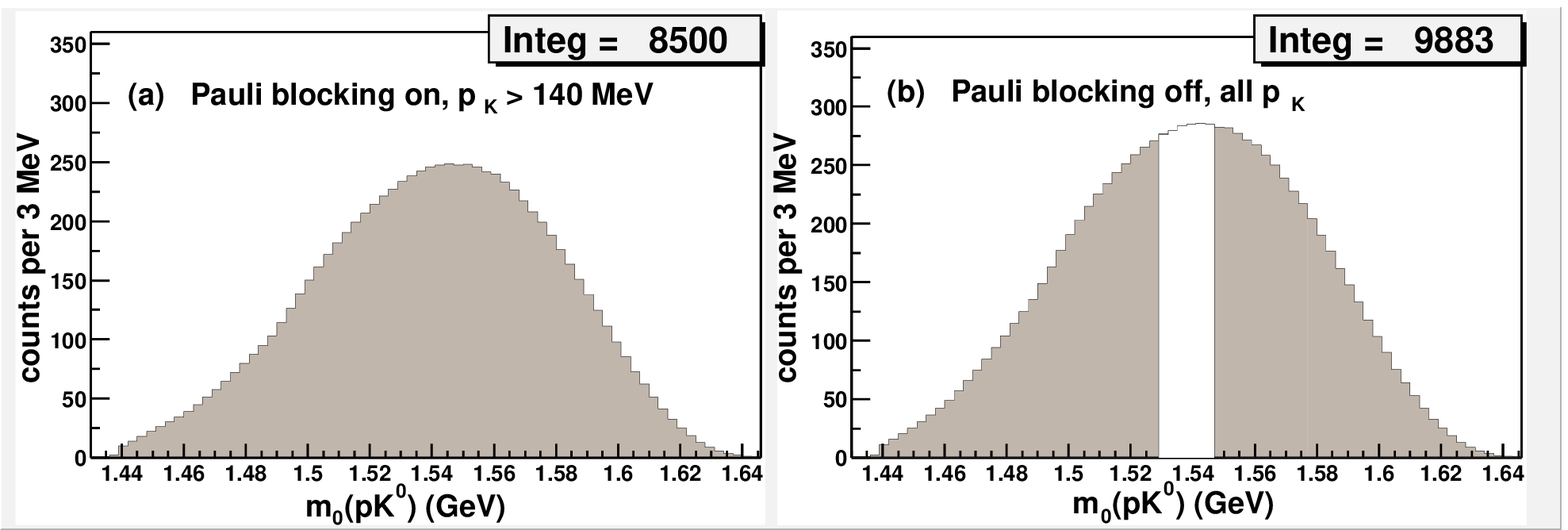}
\caption {(Color online)
The simulated effective mass of the original $pK^0$ system prior 
to any intranuclear rescatterings, $m_0(pK^0)$, for the restricted
fiducial volume $L(K^+) > 520$ mm corresponding to the region of 
throughput measurements (a). The $K^0$ lab momentum is restricted 
to $p_K > 140$ MeV which is the effective threshold for detecting
$K^0_S \rightarrow \pi^+\pi^-$ decays in the scan, and the distribution 
is normalized to the total number of $K^+\mathrm{Xe} \rightarrow K^0_S X$
collisions with $K^0_S \rightarrow \pi^+\pi^-$ decays found by 
the scan in the aforementioned region $L(K^+) > 520$ mm. 
Switching off Pauli blocking in the simulation and lifting the 
cut $p_K > 140$ MeV results in the $m_0(pK^0)$ spectrum that
is shown in (b). The open-white corridor in the latter histogram 
depicts the mass region $1529 < m_0(pK^0) < 1547$ MeV.}
\label{undistorted}
\end{figure}

     For the simulated charge-exchange collisions \under\ on a bound 
neutron in the bubble chamber DIANA, the effective mass of the original 
$pK^0$ system is plotted in Fig.~\ref{undistorted}a. At this stage,
Pauli blocking is still present in the simulation to allow for an
absolute normalization based on the scanning information. The $K^0$
lab momentum is restricted to $p_K > 140$ MeV which is the (effective) 
threshold for detecting the $K^0_S \rightarrow \pi^+\pi^-$ decays in 
the scan. The simulated $m_0(pK^0)$ distribution of 
Fig.~\ref{undistorted}a has been scaled to the total number of 
$K^+\mathrm{Xe} \rightarrow K^0_S X$ collisions with 
$K^0_S \rightarrow \pi^+\pi^-$ decays and any number of detected 
protons, that have been found by the scan 
in the restricted fiducial volume $L(K^+) > 520$ mm ($8500\pm540$
events). Thereby, we obtain the correctly normalized $m_0(pK^0)$
distribution for all events of the charge-exchange reaction
$K^+\mathrm{Xe} \rightarrow K^0_S X$ found by the scan in the
part of the detector fiducial volume where throughput measurements
were made. The next step is to switch off the Pauli suppression and 
lift the selection $p_K > 140$ MeV in the simulation. The resultant 
$m_0(pK^0)$ distribution, that is shown in Fig.~\ref{undistorted}b, 
can be directly used for estimating the ``equivalent" non-resonant
background under the $\Theta^+$ peak. 

     In Eq.~1, we substitute $\Delta m_0 = 18$ MeV and 
$N_\mathrm{bkgd} = 1696\pm108$ events that populate the mass interval
$1529 < m_0(pK^0) < 1547$ MeV of the simulated $m_0(pK^0)$ distribution 
of Fig.~\ref{undistorted}b. The fit of the resonant signal prior to
analysis selections, shown in Fig.~\ref{1dim-all-and-sele}a, returned 
$83.1\pm22.1$ events above the background. This has to be corrected for 
the experimental losses due to secondary interactions of the $K^0$ or 
proton in liquid Xenon, $K^0_S$ mesons decaying too close to the primary 
vertex, and technical reasons for which some events could not be properly
measured. The correction factor for these losses is estimated 
as $1.67\pm0.19$. The $\Theta^+$ signal must also be corrected for the
cuts $p_K > 155$ MeV and $p_p > 165$ MeV and the losses of soft
secondaries, as well as for the experimental selections 
$\Theta_{\pi\pi} < 150^{0}$ and $p_\pi > 75$ MeV for the decays
$K^0_S \rightarrow \pi^+\pi^-$. The corresponding correction factor is 
estimated as 1.43 using a simulation of $\Theta^+$ formation and decay.
And finally, the signal must be corrected for the cut $\tau < 3\tau_0$
on the $K^0_S$ proper lifetime. As a result, in Eq.~1
we have to substitute $N_\mathrm{peak}=208\pm60$ for the 
acceptance-corrected signal of the $\Theta^+$ baryon.

     Finally, for the intrinsic width of the $\Theta^+$ baryon
we obtain $\Gamma = 0.34\pm0.10$ MeV, where the error does not include 
the systematic uncertainties of the simulation procedure. This estimate
of the $\Theta^+$ intrinsic width has been derived assuming that the 
bulk of produced $\Theta^+$ baryons decay upon leaving the nucleus. 
The value of $\Gamma$ obtained in this analysis is consistent with our 
earlier estimates \cite{DIANA-2007,DIANA-2010}, and does not violate 
the upper limits set by BELLE \cite{BELLE} and by the E19 experiment 
at J-PARC \cite{Shirotori}.

\section{Summary and conclusions}

     We have analyzed the DIANA data on the 
charge-exchange reaction \charex\ using increased statistics and 
modified selections. The distribution of the $pK^0$ effective mass 
shows a prominent enhancement at 1538 MeV whose width is consistent
with being entirely due to the experimental resolution. Applying the same 
selections as in our previous analysis \cite{DIANA-2003}, we find that 
this narrow enhancement has increased proportionally to the increase 
of the data sample. A corresponding enhancement at 
\dimass$\simeq1538$ MeV and $p(K^+)\simeq 490$ MeV, formed by nearly 80 
events above the background, is observed in the scatter plot in the
variables \dimass\ and $p(K^+)$. Relying on a simulation of $K^+$Xe
collisions that includes the development of the intranuclear cascade,
we have shown that the observed signal is not a spurious structure 
created by the selections. Under the additional kinematic selections 
based on the simulation, the statistical significance of the signal
reaches 5.5 standard deviations. We interpret 
our observations as strong evidence for formation of a pentaquark 
baryon with positive strangeness, $\Theta^+(uudd\bar{s})$, in the 
charge-exchange reaction \under\ on a bound neutron. The mass of the 
$\Theta^+$ baryon has been measured as $m(\Theta^+) = 1538\pm2$ MeV. 
Using the ratio between the numbers of resonant and non-resonant 
charge-exchange events in the peak region, the intrinsic width of 
this baryon resonance has been determined as
$\Gamma(\Theta^+) = 0.34\pm0.10$ MeV.
The results reported in this paper confirm our earlier observations 
based on part of the present statistics of the charge-exchange 
reaction \charex\ \cite{DIANA-2003,DIANA-2007,DIANA-2010}.

     We wish to thank M.~B. Zhalov for communicating his results on
Fermi momentum in the Xe nucleus. We also thank Ya.~I. Azimov, 
L.~N. Bogdanova, D.~I. Diakonov, and I.~I. Strakovsky for useful 
comments. This work is supported by the Russian Foundation for 
Basic Research (grant 10-02-01487).

\end{document}